\documentclass[sigconf]{acmart}

\renewcommand\footnotetextcopyrightpermission[1]{}

\makeatletter
\renewcommand\acmConference[3]{}
\renewcommand\acmBooktitle{}
\renewcommand\@acmBooktitle{}
\makeatother
\usepackage{longtable}
\usepackage{booktabs}
\usepackage{tabularx}
\AtBeginDocument{%
  }

\begin{document}
\pagestyle{plain}

\title{Predicting Juror Predisposition Using Machine Learning: A Comparative
Study of Human and Algorithmic Jury Selection}

\author{Ashwin Murthy}
\affiliation{%
\institution{Amazon}
\city{Sunnyvale}
\state{California}
\country{USA}}
\email{ashwinmurthy@amazon.com}

\author{Ramesh  Krishnamaneni}
\affiliation{%
\institution{IBM}
\city{Dallas}
\state{Texas}
\country{USA}}
\email{rameshk@ibm.com}

\author{Sean Chacon}
\affiliation{%
\institution{Caremont Graduate University}
\city{Claremont}
\state{California}
\country{USA}}
\email{sean.chacon@cgu.edu}

\author{Kelsey Carlson}
\affiliation{%
\institution{Independent Researcher}
\city{Greensboro}
\state{North Carolina}
\country{USA}}
\email{dr.kelseycarlson@gmail.com}

\author{Ranjita Naik}
\affiliation{%
\institution{Georgia Institute of Technology}
\city{Atlanta}
\state{Georgia}
\country{USA}}
\email{rnaik36@gatech.edu}


\begin{abstract}
Prior studies on the effectiveness of professional jury consultants in predicting juror proclivities have yielded mixed results, and few have rigorously evaluated consultant performance against chance under controlled conditions. This study addresses that gap by empirically assessing whether jury consultants can reliably predict juror predispositions beyond chance levels and whether supervised machine-learning (ML) models can outperform consultant predictions.

Using data from N mock jurors who completed pre-trial attitudinal questionnaires and rendered verdicts in a standardized wrongful-termination case, we compared predictions made by professional jury consultants with those generated by Random Forest (RF) and k-Nearest Neighbors (KNN) classifiers. Model and consultant predictions were evaluated on a held-out test set using paired statistical tests and nonparametric bootstrap procedures.

We find that supervised ML models significantly outperform professional jury consultants under identical informational constraints, while offering greater transparency, replicability, and auditability. These results provide an empirical benchmark for evaluating human judgment in jury selection and inform ongoing debates about the role of data-driven decision support in legal contexts. To support reproducibility and auditability, all code and data will be made publicly available upon publication.
\end{abstract}





\maketitle

\section{Introduction}
\begin{figure}[t]
    \centering
    \includegraphics[width=0.8\linewidth]{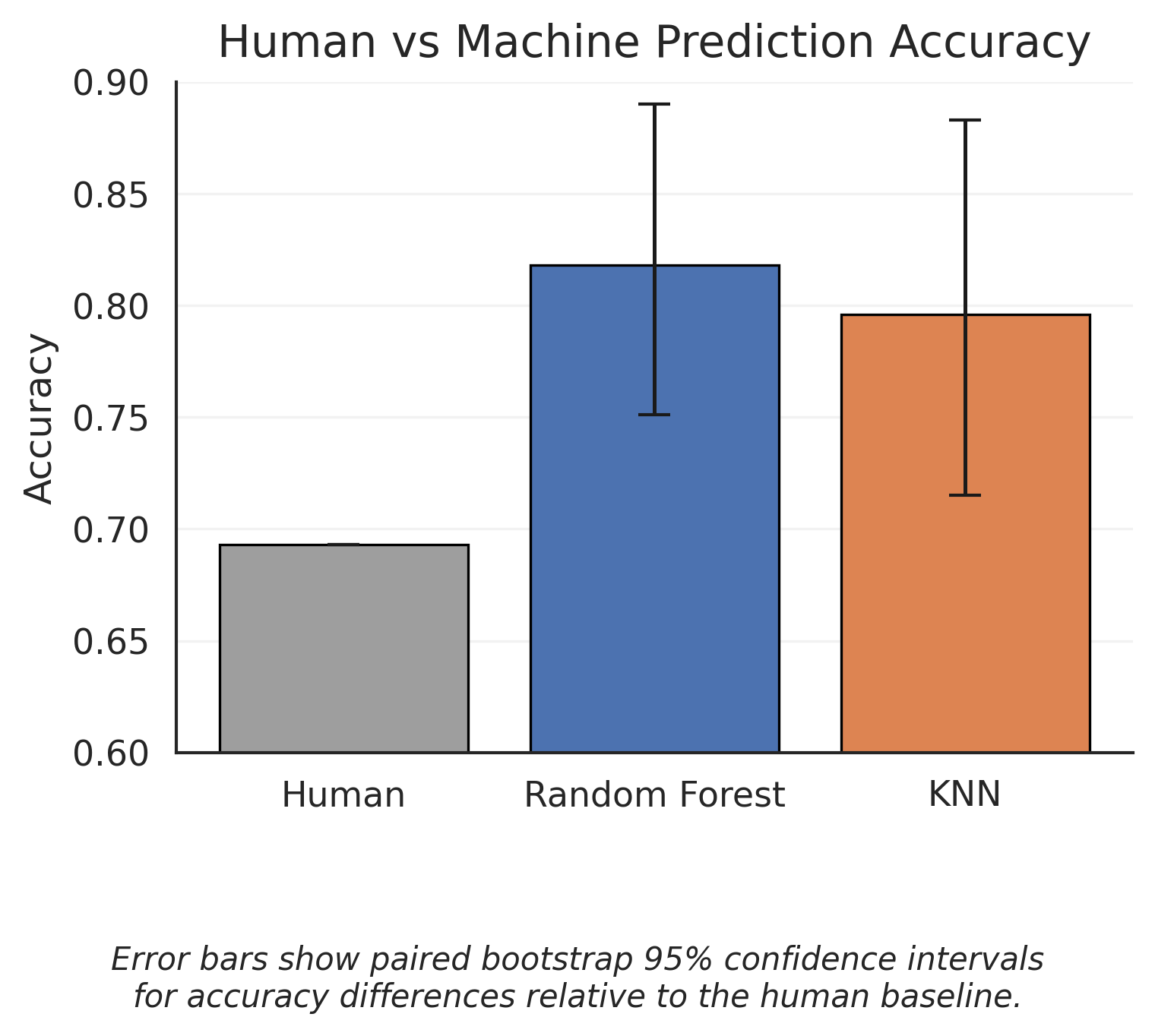}
    \caption{Comparison of predictive accuracy between professional jury consultants (majority vote) and machine-learning models. Error bars denote 95\% paired bootstrap confidence intervals relative to the human baseline.}
    \label{fig:teaser}
\end{figure}

The Sixth and Seventh Amendments protect the right to an impartial jury in criminal and civil proceedings. Voir dire is the primary mechanism through which that right is safeguarded: attorneys question prospective jurors to determine whether their beliefs or experiences might affect their ability to judge a case fairly. In practice, however, voir dire unfolds under severe informational constraints, brief questioning, short questionnaires, and rapid judgments formed under pressure.

To mitigate these constraints, litigants, particularly in high-stakes trials, often retain professional jury consultants, sometimes referred to as practitioners of ``scientific jury selection.'' These consultants typically analyze juror questionnaires, demographic information, and qualitative impressions to advise attorneys on juror strikes and trial strategy. Despite the growing institutionalization of this practice, several structural limitations persist.

First, cost and access sharply constrain the reach of jury consulting. Trial consulting services are expensive and are therefore disproportionately available to well-resourced litigants. Prior scholarship notes that comprehensive consulting engagements can cost tens of thousands of dollars or more, effectively placing such services beyond the reach of public defenders, legal aid organizations, and many small firms \citep{Griffith2007TrialConsultants, WillmottOostinga2017ScientificJurySelection}. This economic barrier raises concerns that purportedly ``scientific'' jury selection tools may exacerbate existing inequalities in the justice system rather than mitigate them.

Second, trial consultant judgments are largely grounded in professional experience and qualitative interpretation rather than standardized, empirically validated predictive models. As a profession, trial consulting lacks uniform credentialing requirements, specialized training standards, or the use of standardized assessments when advising on jury selection decisions \citep{Lecci2004}. This absence of formal regulation and methodological standardization contributes to an opaque decision-making process in which consultant recommendations are difficult to independently evaluate, replicate, or compare across cases. As Lecci and colleagues argue, the esoteric and idiosyncratic nature of trial consulting places the burden on the profession to empirically demonstrate its efficacy, an obligation that can only be met through systematic, quantitative validation rather than post hoc claims of success.

Finally, and most critically, there remains a lack of robust empirical validation of consultant accuracy. While consultants are widely used in practice, few studies have rigorously tested whether their predictions of juror proclivities outperform chance under controlled conditions, or whether they generalize across cases and populations. Existing empirical findings are mixed, with some studies questioning whether consultant judgments meaningfully exceed baseline prediction strategies \citep{Heise2025JuryConsultants, Salerno2021VoirDire, Seltzer2006}. As a result, the evidentiary foundation for scientific jury selection remains limited relative to its influence in modern trial practice.

These limitations raise fundamental questions about the reliability, fairness, and accountability of prevailing approaches to jury selection, motivating the need for systematic, empirically grounded alternatives capable of being evaluated, audited, and replicated.

Advances in artificial intelligence (AI) and machine learning (ML) offer one potential response to these challenges. ML models have demonstrated strong performance in predicting judicial outcomes, identifying latent legal patterns, and capturing nonlinear relationships that human judgment struggles to synthesize \citep{Aletras2016ECHR, Medvedeva2020ECHR}. Research on human–algorithm interaction further suggests that algorithmic predictions are often viewed as more consistent and objective once users are exposed to their performance \citep{Logg2019AlgorithmAppreciation}.

At the same time, applying ML to juror assessment raises significant constitutional and ethical concerns. Equal-protection doctrine, including \emph{Batson v. Kentucky} (1986), prohibits evaluating prospective jurors on the basis of race or sex. Yet research in algorithmic fairness shows that predictive models can encode discriminatory proxy signals even when sensitive features are removed \citep{Dressel2018Recidivism}. Scholars in computational law argue that bias in ML systems is often multidimensional, requiring careful auditing of error patterns and decision behavior to ensure legal and normative legitimacy \citep{Roy2023Discrimination, Mitchell2019ModelCards}. Accordingly, ML-based approaches to voir dire must be evaluated not only for predictive performance but also for transparency, replicability, and error structure.

Despite these developments, juror-level prediction remains underexplored. Psycholegal research has established that pretrial attitudes reliably forecast verdict direction across legal contexts \citep{Lundrigan2016Verdicts, Salerno2021VoirDire}, yet few studies have built predictive models using such data, and almost none have directly compared algorithmic predictions to those made by professional jury consultants under identical informational constraints.

In this study, we conduct a controlled, juror-level comparison of professional jury consultants and supervised machine-learning models using the same pretrial questionnaire information. Figure~\ref{fig:teaser} provides a high-level preview of the central finding: under identical conditions, ML models achieve substantially higher predictive accuracy than consultant majority votes. The remainder of the paper systematically evaluates the robustness, statistical significance, and structure of these performance differences.

Accordingly, we investigate the following research questions:
\begin{itemize}
    \item \textbf{RQ1:} Do professional jury consultants predict juror verdict leanings at rates that exceed chance?
    \item \textbf{RQ2:} Do supervised machine-learning models predict juror verdict leanings more accurately than professional jury consultants when provided with identical pretrial questionnaire information?
    \item \textbf{RQ3:} Do machine-learning models exhibit more consistent and systematic error patterns than professional jury consultants on individual jurors?
    \item \textbf{RQ4:} Are attitudinal factors more influential than demographic characteristics in predicting juror verdict leanings?
\end{itemize}

\section{Related Work}

Our work draws on four intersecting bodies of scholarship: (1) empirical research on juror decision-making and voir dire, (2) critiques and evaluations of scientific jury selection and trial consulting, (3) comparisons of human and algorithmic judgment, and (4) machine learning in legal decision-making and fairness-sensitive contexts.

\subsection{Juror Decision-Making and Voir Dire}

A large body of psycholegal research has examined how jurors form decisions and the role of pre-trial attitudes in shaping verdicts. Decades of empirical work demonstrate that juror beliefs, values, and prior experiences systematically influence trial outcomes, even before evidence is presented \citep{Devine2001JuryDecision, Curley2022JurorBias, Castrellon2023JurorBias}. Attitudinal factors such as beliefs about fairness, responsibility, and authority have been shown to predict verdict direction across civil and criminal contexts.

Within this literature, voir dire has been studied as a mechanism for identifying and managing juror bias. Experimental work suggests that expanded voir dire can affect juror decisions, though its effectiveness varies by case type and judicial intervention \citep{Salerno2021VoirDire}. Related studies have shown that pre-trial attitudes, combined with the applicable standard of proof, can meaningfully predict verdicts, reinforcing the relevance of questionnaire-based juror assessment \citep{Lundrigan2016Verdicts}.

\subsection{Scientific Jury Selection and Trial Consulting}

Professional jury consulting, often framed as ``scientific jury selection,'' applies social-scientific insights to jury selection and trial strategy. Descriptive accounts document how consultants use demographic, attitudinal, and experiential information to advise attorneys \citep{Griffith2007TrialConsultants, WillmottOostinga2017ScientificJurySelection}. However, the empirical evidence supporting consultant effectiveness remains limited and mixed.

Early evaluations questioned whether consultant predictions reliably outperform chance or attorney intuition \citep{Lecci2004, Seltzer2006}. More recent legal scholarship has reiterated this critique, noting the absence of systematic benchmarking, standardized training, or transparent validation practices within the profession \citep{Heise2025JuryConsultants}. Despite widespread use in high-stakes litigation, few studies have directly compared consultant predictions against quantitative models under controlled conditions, leaving core claims of scientific validity largely untested.

\subsection{Human Judgment, Noise, and Expert Prediction}

Research in judgment and decision-making has documented substantial variability and noise in expert assessments across domains. Even trained professionals often struggle to integrate multiple probabilistic cues consistently, leading to unstable or idiosyncratic judgments \citep{Dror2020Bias}. Work in organizational psychology similarly shows that structured statistical models frequently outperform human experts in selection and prediction tasks by reducing noise and enforcing consistent aggregation rules \citep{Highhouse2023Noise, Yu2022ExpertJudgment}.

At the same time, studies of algorithm appreciation suggest that people may prefer algorithmic advice once its consistency and performance are made explicit \citep{Logg2019AlgorithmAppreciation}. These findings motivate empirical comparisons between human and algorithmic judgment, particularly in domains where expert authority has historically gone unchallenged.

\subsection{Machine Learning, Law, and Fairness}

Recent work has applied machine learning to legal prediction tasks, including forecasting judicial decisions using textual and structured data \citep{Aletras2016ECHR, Medvedeva2020ECHR}. While these studies demonstrate the predictive capacity of ML in legal contexts, they also raise concerns about interpretability, bias, and normative legitimacy.

Research on algorithmic fairness emphasizes that predictive accuracy alone is insufficient in high-stakes legal settings. Models may encode proxy discrimination even when sensitive attributes are excluded \citep{Dressel2018Recidivism}, and discrimination can emerge across intersecting demographic dimensions \citep{Roy2023Discrimination}. In response, transparency and accountability frameworks, such as model cards, have been proposed to support responsible evaluation and governance of ML systems \citep{Mitchell2019ModelCards}.

Equal-protection doctrine further constrains the use of predictive tools in jury selection. In \emph{Batson v. Kentucky}, the U.S. Supreme Court prohibited the use of race-based peremptory challenges, underscoring the legal risks of both human and algorithmic decision-making processes that rely on protected characteristics \citep{Batson1986}. These constraints highlight the importance of auditing, benchmarking, and careful governance when introducing data-driven tools into voir dire.

\subsection{Positioning of the Present Work}

Despite extensive research on juror attitudes, trial consulting, and algorithmic prediction, few studies have conducted direct, head-to-head comparisons between professional jury consultants and machine-learning models using identical information and paired evaluation. The present study addresses this gap by empirically benchmarking consultant judgments against supervised ML models under controlled conditions, while explicitly separating predictive performance from normative and legal permissibility.

By situating algorithmic prediction as a transparent, auditable benchmark rather than a replacement for human judgment, this work contributes to ongoing debates about the role of data-driven tools in legally and ethically constrained decision-making contexts.

\section{Methods}

\subsection{Study Design and Data Collection}

We conducted a controlled mock-trial experiment using an online crowdsourcing platform, following established methodologies in empirical legal and psycholegal research (e.g., Salerno et al., 2021). Participants were recruited to serve as mock jurors in a standardized wrongful-termination case.

Each juror (\(N = 410\)) completed a pre-trial questionnaire designed to capture three classes of variables: (1) demographic characteristics (e.g., age, gender, education, employment status), (2) experiential factors (e.g., prior jury service, workplace experience, exposure to discrimination claims), and (3) attitudinal measures assessing beliefs about workplace fairness, discrimination, corporate responsibility, and individual accountability. Attitudinal items were measured using multi-point Likert scales.

Participants then reviewed a standardized case vignette summarizing the relevant facts, legal claims, and arguments presented by both parties. After reviewing the materials, jurors rendered a verdict decision: \emph{plaintiff}, \emph{defense}, or \emph{neither}. To avoid imposing an artificial binary choice that could introduce label noise, jurors were explicitly offered the \emph{neither} option. For predictive modeling, only definitive plaintiff or defense verdicts were retained as ground-truth labels; jurors selecting \emph{neither} were excluded from subsequent analyses.

\subsection{Human Jury Consultant Predictions}

Three professional jury consultants independently reviewed the same juror questionnaires and case materials. Consultants were blind to jurors’ actual verdicts and to one another’s predictions. Based solely on pre-trial information available during voir dire, each consultant classified jurors as either \emph{plaintiff-leaning} or \emph{defense-leaning}. No constraints were imposed on consultants’ reasoning strategies, mirroring real-world jury consulting practice.

Inter-rater reliability among consultants was assessed using Cohen’s~\(\kappa\), yielding moderate agreement (\(\kappa = 0.76\)). Consultant predictions are reported  using majority-vote aggregation.

\subsection{Feature Engineering and Preprocessing}

Questionnaire responses were preprocessed prior to model training. Demographic variables were encoded using one-hot encoding to avoid imposing ordinal assumptions, while Likert-scale attitudinal variables were ordinally encoded to preserve their monotonic structure.

Because several attitudinal measures exhibited moderate correlations, models were evaluated under two feature configurations: (1) a full-feature configuration including all demographic and attitudinal variables, and (2) a random-subspace configuration in which models were trained on randomly sampled subsets of predictors. This design enabled assessment of robustness to feature redundancy and multicollinearity.

All preprocessing steps were fit exclusively on the training data and applied unchanged to the test set.

\subsection{Supervised Machine-Learning Models}

We evaluated two supervised machine-learning models that represent complementary inductive biases and are well-suited to structured questionnaire data: Random Forests (RF) and k-Nearest Neighbors (KNN).

\textbf{Random Forests (RF)} are ensemble bagging methods that aggregate predictions across multiple decision trees trained on bootstrapped samples of the data. By combining recursive partitioning with feature subsampling, RF models can capture non-linear interactions and higher-order dependencies among juror characteristics while remaining relatively robust to multicollinearity and noise. Random Forests are widely used in applied social-science and legal prediction tasks due to their strong performance on tabular data and their ability to provide stable predictions without strong parametric assumptions.

\textbf{k-Nearest Neighbors (KNN)} is a non-parametric, instance-based classifier that assigns labels based on the majority class among the $k$ most similar observations in feature space. Unlike tree-based models, KNN makes minimal assumptions about the global structure of the data and instead relies on local similarity relationships. As such, KNN provides a useful contrast to Random Forests by emphasizing proximity-based reasoning rather than learned decision boundaries.

Together, these models span two distinct predictive paradigms ensemble-based global modeling and local instance-based inference, allowing us to assess whether juror-leaning predictions are driven by stable global patterns or by localized clusters of similar jurors.

Both models were trained using identical feature representations, training procedures, and evaluation protocols, enabling direct and controlled comparison with one another and with human jury consultant predictions.

\subsection{Training and Evaluation Protocol}

The dataset was partitioned into a fixed training set of 273 jurors and an independent held-out test set of 137 jurors, each with 29 features. Model development, including hyperparameter tuning, was conducted exclusively on the training data using 5-fold cross-validation.

Hyperparameters for each model were optimized via grid search within the cross-validation loop. The best-performing configuration was then retrained on the full training set and evaluated once on the held-out test set. This procedure ensured that all reported test-set results reflect strictly out-of-sample performance.

Model performance was evaluated using standard classification metrics, including accuracy, precision, recall, and area under the ROC curve (AUC). To contextualize algorithmic performance, machine-learning predictions were compared directly with jury consultant predictions on the same test-set jurors, enabling head-to-head comparison under identical informational constraints.

\subsection{Limitations and Scope}

The present study evaluates predictive accuracy under controlled experimental conditions and does not claim deployment readiness. While fairness and subgroup analyses are critical for real-world use, they are outside the scope of this work and are identified as priorities for future research.

\subsection{Evaluation Metrics}

Model performance was evaluated on a fixed, independent held-out test set of 137 jurors. We report standard classification metrics including accuracy, precision, recall, and F1-score. Accuracy is emphasized as the primary metric to facilitate direct comparison with binary human consultant judgments.

To quantify uncertainty and avoid reliance on asymptotic assumptions, we estimated confidence intervals using nonparametric bootstrapping. Specifically, the test set was resampled with replacement for 5,000 bootstrap iterations. For each resample, model predictions were recomputed and evaluation metrics were recorded. The resulting empirical distributions were used to estimate mean performance and 95\% confidence intervals.

To enable statistically grounded comparison between machine-learning models and professional jury consultants, we conducted paired analyses on identical test-set jurors. Differences in accuracy between each model and the consultant majority vote were assessed using paired bootstrap resampling, yielding confidence intervals for accuracy deltas. In addition, McNemar’s test was applied to paired predictions to evaluate whether observed differences in error patterns between models and consultants were statistically significant.

All reported confidence intervals and statistical tests were computed exclusively on held-out test data, and no information from the test set was used during model training or hyperparameter tuning.

\section{Results}

\begin{table*}[t]
\centering
\caption{Predictive performance on the held-out test set (137 jurors). $\Delta$Accuracy reflects paired differences relative to the human consultant majority vote, with 95\% bootstrap confidence intervals.}
\label{tab:main_results}
\begin{tabular}{lcccccc}
\toprule
Model & Accuracy & Precision & Recall & F1 & $\Delta$Accuracy vs Human & McNemar $p$ \\
\midrule
Human (Majority Vote) & 0.693 & 0.720 & 0.756 & 0.738 & -- & -- \\
Random Forest & 0.818 & 0.827 & 0.859 & 0.843 & 0.123 [0.058, 0.197] & 0.001 \\
KNN & 0.796 & 0.784 & 0.885 & 0.831 & 0.101 [0.022, 0.190] & 0.026 \\
\bottomrule
\end{tabular}
\end{table*}

\begin{figure*}[t]
    \centering
    \includegraphics[width=0.85\textwidth]{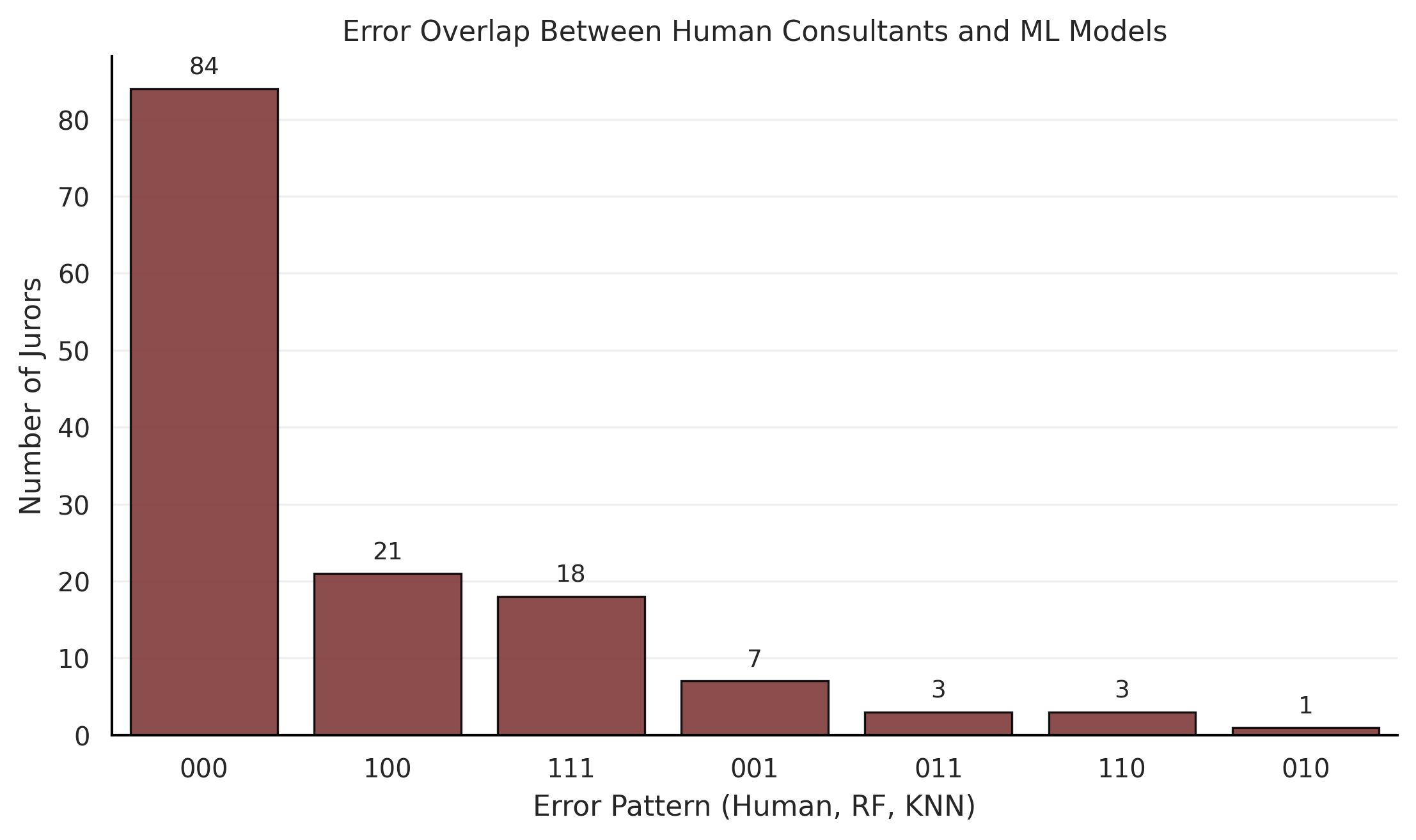}
    \caption{Error overlap patterns across human consultants, Random Forest, and k-Nearest Neighbors models. Each bar represents the number of jurors corresponding to a specific error pattern, where 1 indicates an incorrect prediction and 0 indicates a correct prediction (ordered as Human, RF, KNN).}
    \label{fig:error_overlap}
\end{figure*}

\subsection{Overall Predictive Performance}

Table~\ref{tab:main_results} reports predictive performance for the human jury consultant majority vote, Random Forest (RF), and k-Nearest Neighbors (KNN) models on the held-out test set of 137 jurors. Both machine-learning models substantially outperformed the human consultant baseline in terms of accuracy, precision, recall, and F1-score.

The human consultant majority vote achieved an accuracy of 0.693. In contrast, the Random Forest model achieved an accuracy of 0.818, while the KNN model achieved an accuracy of 0.796. Improvements were consistent across precision and recall, indicating that gains were not driven by asymmetric error tradeoffs.

\subsection{Comparison with Human Jury Consultants}

To directly compare model performance against professional jury consultants, we evaluated paired accuracy differences on identical test-set jurors. Random Forest improved accuracy by 12.3 percentage points relative to the consultant majority vote, with a 95\% paired bootstrap confidence interval of [0.058, 0.197]. KNN achieved a 10.1 percentage point improvement, with a corresponding confidence interval of [0.022, 0.190].

Importantly, confidence intervals for both models exclude zero, indicating that observed performance gains are unlikely to be attributable to sampling variability alone.

\subsection{Statistical Significance of Error Differences}

We further assessed whether models and consultants differed in their error patterns using McNemar’s test. Random Forest predictions differed significantly from consultant predictions (\(p = 0.001\)), with substantially more cases in which the model was correct and the consultant was incorrect than vice versa (\(b = 21, c = 4\)). KNN predictions also differed significantly from consultants (\(p = 0.026\)), though with a smaller margin (\(b = 24, c = 10\)).

Together, these results indicate that machine-learning models not only achieve higher overall accuracy than professional jury consultants but also make systematically different and more frequently correct judgments on individual jurors under identical informational constraints.

\subsection{Confusion Matrix Analysis}

\begin{figure*}[t]
    \centering
    \includegraphics[width=0.95\textwidth]{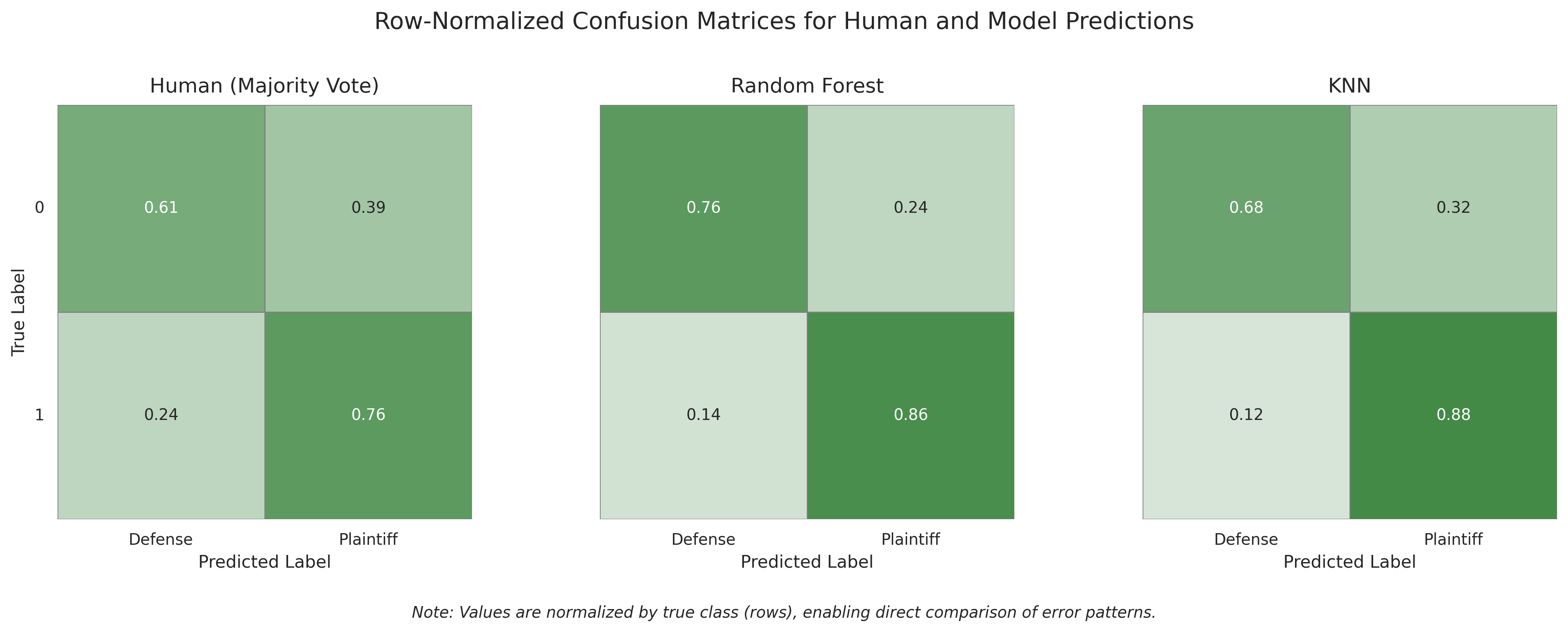}
    \caption{Row-normalized confusion matrices comparing the human jury consultant majority vote, Random Forest, and k-Nearest Neighbors predictions on the held-out test set. Values indicate the proportion of jurors within each true class assigned to each predicted class, enabling direct comparison of error patterns across decision systems.}
    \label{fig:confusion_matrices}
\end{figure*}

Figure~\ref{fig:confusion_matrices} presents row-normalized confusion matrices for the human jury consultant majority vote, Random Forest, and k-Nearest Neighbors models. Both machine-learning models reduce misclassification rates for plaintiff- and defense-leaning jurors relative to consultants. Random Forest exhibits the most balanced improvements across outcome classes, while KNN shows particularly strong performance on plaintiff-leaning jurors. These results indicate that model gains reflect broad reductions in error rather than asymmetric trade-offs between false positives and false negatives.

\subsection{Error Structure and Predictive Consistency}

To further characterize differences between human and algorithmic judgment, we examined patterns of agreement and disagreement on individual jurors. Figure~\ref{fig:error_overlap} summarizes error overlap across the human consultant majority vote, Random Forest, and k-Nearest Neighbors models.

The most common pattern corresponds to unanimous agreement across predictors, indicating broad consensus on a majority of jurors. Notably, a substantial number of cases reflect instances in which both machine-learning models correctly predicted juror leanings while the consultant majority vote erred. In contrast, cases in which the consultant was correct and both models were incorrect were comparatively rare. A smaller subset of jurors was misclassified by all predictors, suggesting the presence of inherently ambiguous cases rather than random noise.

These error patterns complement the McNemar test results, indicating that the observed performance gains reflect systematic differences in predictive consistency rather than isolated improvements in aggregate accuracy.

\subsection{Attitudinal Versus Demographic Predictors}

To assess the relative influence of attitudinal and demographic factors, we conducted both performance-based ablation analyses and feature-importance inspections. Models trained exclusively on attitudinal variables achieved identical predictive performance to models that additionally included demographic attributes such as race and gender (Table~\ref{tab:ablation}), indicating that demographic features do not materially contribute to predictive accuracy in this setting.

\begin{figure}[t]
    \centering
    \includegraphics[width=\linewidth]{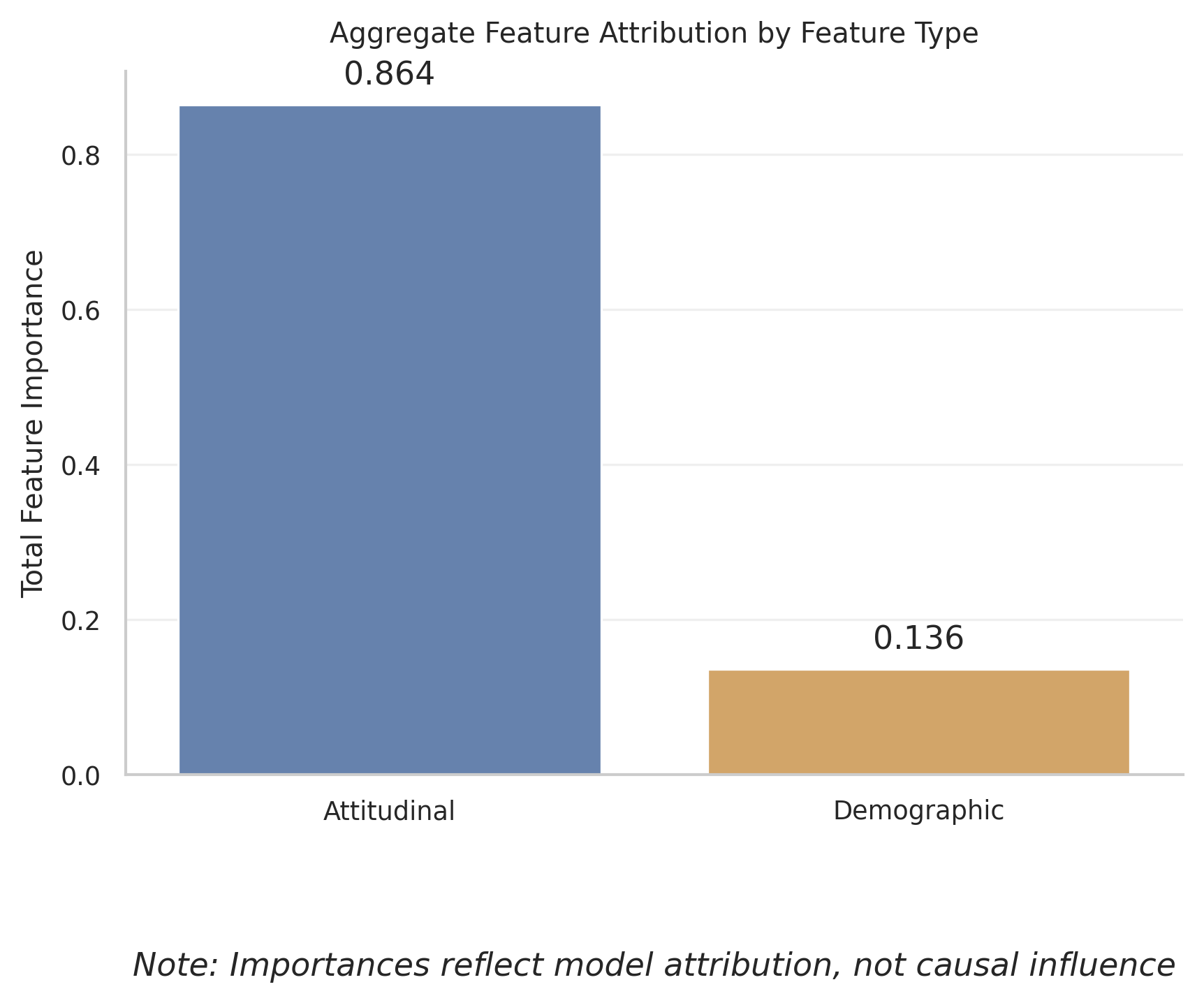}
    \caption{Aggregate feature importance by feature type for the Random Forest model. Attitudinal variables account for the majority of predictive importance, while demographic attributes contribute substantially less.}
    \label{fig:aggregate_importance}
\end{figure}

Figure~\ref{fig:aggregate_importance} summarizes aggregate feature importance by feature type for the Random Forest model. Attitudinal variables account for the overwhelming majority of total importance, whereas demographic attributes contribute substantially less to the model’s predictions.

\begin{figure*}[t]
    \centering
    \includegraphics[width=1.0\textwidth]{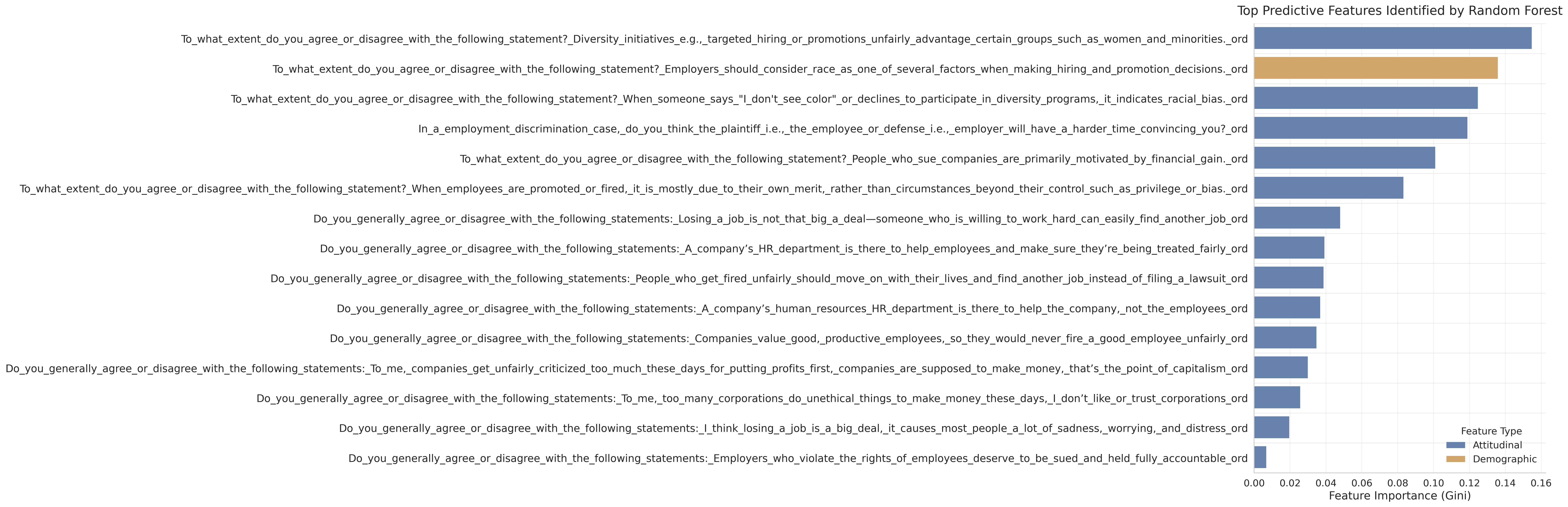}
    \caption{Top predictive features identified by the Random Forest model, ranked by Gini importance. The most influential predictors correspond to jurors’ attitudinal beliefs about workplace fairness, discrimination, and responsibility attribution, while demographic attributes exhibit comparatively low importance.}
    \label{fig:feature_importance}
\end{figure*}

Inspection of individual feature importances further reveals that the most influential predictors correspond to jurors’ beliefs about workplace fairness, discrimination, and responsibility attribution (Figure~\ref{fig:feature_importance}). Demographic variables, when included, exhibit comparatively low importance and do not appear among the most predictive features.

Taken together, these findings suggest that predictive signals arise primarily from jurors’ expressed attitudes rather than demographic characteristics, supporting the use of attitudinally grounded models under legally constrained jury-selection contexts.

\paragraph{Robustness to Feature Redundancy.}
To assess robustness to correlated predictors and multicollinearity, we conducted a random-subspace analysis in which models were trained on randomly sampled subsets of attitudinal features while holding the training and test sets fixed. Across both Random Forest and KNN models, predictive performance degraded gradually as the number of available features decreased, without sharp drops indicative of reliance on a narrow feature subset. Even when trained on only 50\% of available predictors, both models retained substantial predictive accuracy. Performance variability also decreased as more features were included, indicating increased stability. Full results are reported in Appendix~\ref{app:subspace}. These findings suggest that model performance reflects distributed attitudinal signal rather than dependence on a small number of correlated variables.

\paragraph{Sensitivity to Demographic Features.}
All primary results exclude protected demographic attributes such as race and gender. As a robustness check, we re-estimated all models including these features while holding the training, evaluation, and comparison procedures constant. Inclusion of demographic attributes did not materially change predictive performance, relative rankings between models and human consultants, or statistical significance (Appendix~X). These findings indicate that the observed performance gains are not driven by the explicit inclusion of protected characteristics.

\begin{table*}[t]
\centering
\caption{Predictive performance with and without demographic features.}
\label{tab:ablation}
\begin{tabular}{lcc}
\toprule
Model & Attitudinal Only & Attitudinal + Demographic \\
\midrule
Random Forest (Accuracy) & 0.818 & 0.818 \\
KNN (Accuracy) & 0.796 & 0.796 \\
\bottomrule
\end{tabular}
\end{table*}

\section{Discussion}

This study provides a controlled, empirical comparison between professional jury consultants and supervised machine-learning models tasked with predicting juror verdict leanings from pre-trial questionnaire data. Under identical informational constraints and evaluated on the same held-out jurors, both Random Forest and k-Nearest Neighbors models significantly outperformed the consultant majority vote. These results have implications for the empirical foundations of scientific jury selection, the role of data-driven tools in trial preparation, and the ethical governance of predictive technologies in legal settings.

\subsection{Interpreting the Performance Gap}

The observed performance gap between machine-learning models and professional jury consultants suggests that juror questionnaire data contain systematic predictive structure that is not fully exploited by human judgment alone. Ensemble-based and instance-based models appear better able to aggregate weak but collectively informative signals across multiple attitudinal and experiential variables, yielding more consistent predictions across jurors.

Importantly, these gains are not marginal. Paired bootstrap confidence intervals for accuracy differences exclude zero for both models, and McNemar tests indicate statistically significant differences in error patterns. This suggests that models are not merely optimizing different tradeoffs but are correctly classifying cases in which consultants systematically err.

These findings align with broader evidence that humans struggle to integrate multiple probabilistic cues consistently, particularly under time pressure and without formal aggregation rules. In contrast, machine-learning models apply stable decision rules across cases, reducing variance introduced by subjective interpretation.

\subsection{Implications for ``Scientific'' Jury Selection}

Professional jury consulting is often described as a form of “scientific jury selection,” yet the field has historically lacked large-scale, controlled validation of predictive accuracy and standardized evaluation frameworks. While our results indicate that professional consultants perform better than chance, the absence of systematic benchmarking, replication, and agreed-upon performance standards raises questions about whether current practices merit the label “scientific.” These findings underscore the need for greater empirical scrutiny to assess when, and under what conditions, consultant judgments can be reliably validated.

This is not a critique of consultant expertise per se. Jury consultants bring contextual knowledge, courtroom experience, and qualitative judgment that are not captured by questionnaire data alone. Rather, the present findings suggest that claims of scientific rigor would benefit from stronger empirical grounding and transparent validation practices. Viewed in this light, quantitative models may serve as a complementary tool—introducing structure, consistency, and testability into a largely subjective process—and as a practical alternative in settings where access to human expertise is limited by cost or resource constraints.

\subsection{Decision Support, Not Automation}

We do not argue that machine-learning models should replace human judgment in jury selection. Rather, our results support a framing of ML as a \emph{decision-support and auditing tool}. Predictive models can serve as external benchmarks against which human judgments are evaluated, helping identify blind spots, inconsistencies, or overconfidence in consultant recommendations.

Such a role is particularly important given the high stakes of jury selection and the constitutional constraints governing it. By providing reproducible and auditable predictions, ML systems may improve accountability without displacing human legal reasoning or strategic decision-making.

\subsection{Fairness, Law, and Constitutional Constraints}

Improved predictive accuracy does not imply normative legitimacy. Jury selection in the United States is governed by constitutional protections, including equal-protection doctrine and the prohibition against race- and sex-based exclusion articulated in \emph{Batson v. Kentucky}. Predictive models trained on juror data may inadvertently encode proxy signals for protected characteristics, even when sensitive attributes are excluded.

While the present study does not conduct subgroup or fairness audits, these results underscore the need for such analyses before any real-world deployment. Performance advantages alone cannot justify the use of predictive systems in contexts where discriminatory impact is legally and ethically impermissible. Future work must therefore integrate fairness auditing, race-sensitivity testing, and legal compliance checks as first-class components of evaluation.

\subsection{Ethical and Legal Considerations}

Any application of machine-learning techniques to jury selection operates within a stringent constitutional and ethical framework. Equal-protection doctrine including the prohibition against race- and sex-based exclusion articulated in \emph{Batson v. Kentucky} places clear limits on how juror information may be used in practice. At the same time, prior research has shown that predictive models can encode discriminatory proxy signals even when protected attributes are excluded from training data.

The present study does not conduct subgroup performance analyses, race-sensitivity tests, or fairness audits. As such, it does not claim that the observed predictive gains satisfy constitutional or ethical requirements. Rather, the findings should be interpreted as evidence about comparative predictive performance under controlled conditions, not as validation of legally permissible deployment.

These constraints underscore the importance of separating performance benchmarking from normative legitimacy. Any real-world use of predictive tools in voir dire would require rigorous fairness auditing, sensitivity analyses, and legal review to assess disparate impact, proxy discrimination, and compliance with constitutional standards. Removing protected attributes may be a necessary step toward responsible model design, but it is not sufficient on its own to guarantee fairness or legality.

\subsection{Access and Institutional Equity}

These findings also have practical implications for access and institutional equity in jury selection. Traditional jury consulting services are costly and disproportionately available to well-resourced litigants, potentially amplifying existing inequalities in trial preparation. In contrast, transparent and standardized machine-learning–based decision-support tools could lower barriers to empirically grounded juror assessment, making such analyses more accessible to public defenders, small firms, and courts.

Importantly, any such use should be framed as advisory rather than determinative. Predictive tools may help flag attitudinal patterns that human evaluators overlook, while remaining subject to human judgment and legal oversight. Existing frameworks from computational law and responsible AI provide pathways for courts to regulate and audit these systems, ensuring that algorithmic assistance enhances rather than undermines procedural fairness.

Notably, including race and gender as features did not materially alter model performance, suggesting that the observed predictive gains are primarily associated with attitudinal information rather than the explicit use of demographic attributes.

\section{Limitations and Future Directions}

Several limitations warrant caution. First, the study relies on a mock-trial design with jurors recruited online, which may limit generalizability to real-world venire members. While online samples are widely used in empirical legal research, they may differ from courtroom jurors in terms of attentiveness, experience, and deliberative context.

Second, the analysis is restricted to a single civil wrongful-termination case. Juror attitudes and consultant strategies may vary across legal domains, evidentiary structures, and stakes. Extending this framework to additional civil and criminal contexts would help assess the stability of attitudinal predictors and comparative performance across case types.

Third, the models and consultant evaluations rely exclusively on structured questionnaire data. In practice, jury selection often incorporates additional signals, including open-ended responses, oral voir dire, and qualitative impressions. Integrating richer or multimodal inputs—such as free-text explanations, audio from voir dire, or interactional cues—may further improve predictive performance and warrants future investigation.

Finally, the present study focuses on predictive accuracy under controlled conditions and does not evaluate fairness, calibration, or downstream legal impacts. Although performance benchmarking is a necessary first step, any real-world application would require rigorous fairness auditing, race-sensitivity testing, and legal review prior to deployment.

Future research should also examine how machine-learning–based decision-support tools influence attorney judgment in practice, how such tools might be regulated by courts, and whether algorithmic insights can be incorporated into structured voir dire procedures to improve transparency and accountability without undermining constitutional protections.

\section{Conclusion}

This study provides a controlled empirical comparison between professional jury consultants and supervised machine-learning models tasked with predicting juror verdict leanings from pre-trial questionnaire data. Under identical informational constraints, Random Forest and k-Nearest Neighbors models achieved significantly higher predictive accuracy than the consultant majority vote.

These findings challenge the assumption that professional jury consultants reliably outperform simple, transparent predictive models and underscore the importance of empirical benchmarking in a domain that has long claimed scientific grounding. At the same time, predictive accuracy alone does not establish normative legitimacy. Machine-learning systems are not a panacea and must be governed carefully, particularly in light of constitutional constraints and ethical concerns surrounding jury selection.

Rather than advocating automation, this work highlights the potential role of machine-learning models as auditable, reproducible benchmarks and decision-support tools for evaluating human judgment in voir dire. Used responsibly, such benchmarks may contribute to more empirically grounded and accountable trial practices while preserving the central role of human legal decision-making.


\section{Ethical Considerations in Mock Juror Recruitment}
Mock jurors were recruited via an online platform commonly used for jury research. Participants were compensated in accordance with applicable minimum wage, time, and labor laws, and all other relevant legal and ethical requirements in the jurisdictions of both the participants and the authors were followed.

\section*{Generative AI Disclosure Statement}

The authors used generative AI tools to assist with code structuring, data analysis workflows, and visualization formatting. All experimental design decisions, model implementations, statistical analyses, and interpretations of results were performed by the authors. Generative AI tools did not autonomously generate research hypotheses, conduct analyses, or determine conclusions.

\begin{acks}
We thank the participating jury consultants and volunteer jurors for their time and contributions to this study. This research received no external funding. The authors declare no conflicts of interest.
\end{acks}


\let\cleardoublepage\clearpage

\appendix
\onecolumn

\section{Feature Importance Details}
\label{app:feature_importance}

Table~\ref{tab:rf_feature_importance_full} reports the top RF feature-importance scores used to assess the relative influence of attitudinal and demographic predictors (RQ3). Importance values are computed using the Gini-importance criterion.

\begin{table*}[t]
\centering
\caption{Top RF feature-importance scores (rounded to four decimal places).}
\label{tab:rf_feature_importance_full}
\begin{tabularx}{\textwidth}{X l r}
\toprule
\textbf{Feature} & \textbf{Type} & \textbf{Importance} \\
\midrule
To what extent do you agree or disagree with the following statement? Diversity initiatives unfairly advantage certain groups & Attitudinal & 0.1549 \\
Employers should consider race as one of several factors in hiring and promotion decisions & Demographic & 0.1359 \\
When someone says ``I don't see color,'' it indicates racial bias & Attitudinal & 0.1248 \\
In an employment discrimination case, which side will have a harder time convincing you? & Attitudinal & 0.1190 \\
People who sue companies are primarily motivated by financial gain & Attitudinal & 0.1010 \\
Employment outcomes are mostly due to merit rather than bias or privilege & Attitudinal & 0.0834 \\
Losing a job is not a big deal for someone willing to work hard & Attitudinal & 0.0480 \\
A company’s HR department is there to help employees & Attitudinal & 0.0393 \\
People fired unfairly should move on instead of filing lawsuits & Attitudinal & 0.0388 \\
A company’s HR department exists to protect the company, not employees & Attitudinal & 0.0369 \\
Companies would never fire a good employee unfairly & Attitudinal & 0.0348 \\
Corporations are unfairly criticized for prioritizing profits & Attitudinal & 0.0300 \\
Too many corporations behave unethically to make money & Attitudinal & 0.0258 \\
Losing a job causes significant distress for most people & Attitudinal & 0.0197 \\

\bottomrule
\end{tabularx}
\end{table*}

\section{Learning Curves and Sample-Size Sensitivity}
As a robustness check, we examined learning curves for the RF and KNN models by varying the size of the training set while evaluating performance using cross-validation. Both models exhibit near-perfect training accuracy across sample sizes, reflecting high model capacity, while cross-validated accuracy improves monotonically as additional data is incorporated. This pattern indicates that the performance gains reported in the main analysis are not driven by overfitting to a small training subset, and that additional data would likely further improve generalization. RF  converges faster than KNN, consistent with ensemble methods’ ability to stabilize predictions in high-dimensional settings.

\begin{table}[t]
\centering
\caption{Random-subspace robustness analysis for RF and KNN models. Each configuration reports mean test accuracy across 50 randomly sampled feature subsets.}
\label{tab:subspace_results}
\begin{tabular}{lcccc}
\toprule
Model & Subspace & \# Features & Mean Acc. & Std. Dev. \\
\midrule
Random Forest & 50\% & 8  & 0.714 & 0.036 \\
Random Forest & 70\% & 12 & 0.768 & 0.024 \\
Random Forest & 90\% & 15 & 0.802 & 0.013 \\
KNN           & 50\% & 8  & 0.723 & 0.054 \\
KNN           & 70\% & 12 & 0.779 & 0.028 \\
KNN           & 90\% & 15 & 0.796 & 0.016 \\
\bottomrule
\end{tabular}
\end{table}

\begin{figure}[t]
    \centering
    \includegraphics[width=0.8\linewidth]{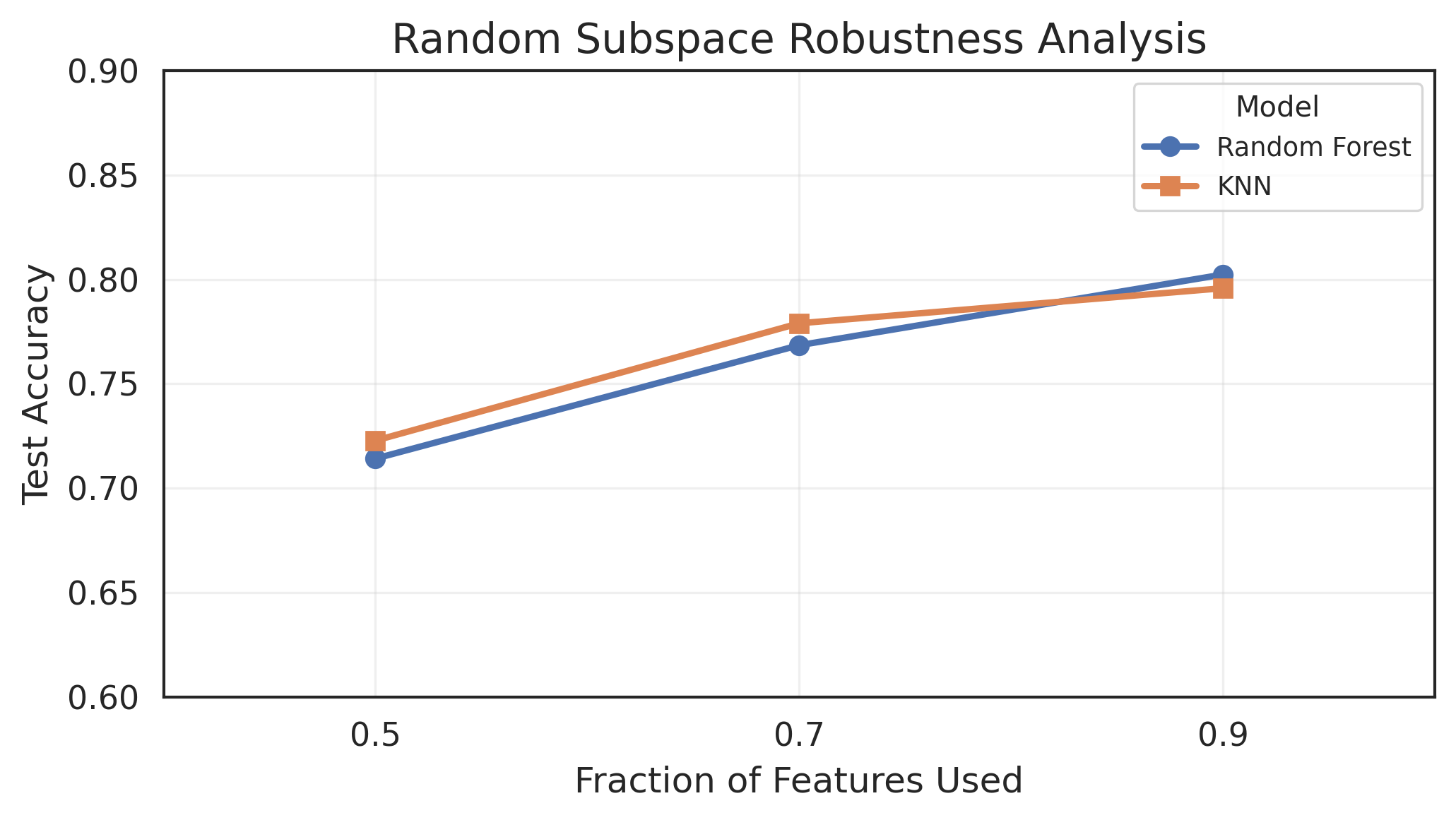}
    \caption{Test accuracy as a function of the fraction of attitudinal features used for RF and KNN models. Accuracy degrades gradually as fewer predictors are available, with no abrupt collapse, indicating robustness to feature redundancy and multicollinearity.}
    \label{fig:subspace_accuracy}
\end{figure}

\begin{figure}[t]
    \centering
    \includegraphics[width=\linewidth]{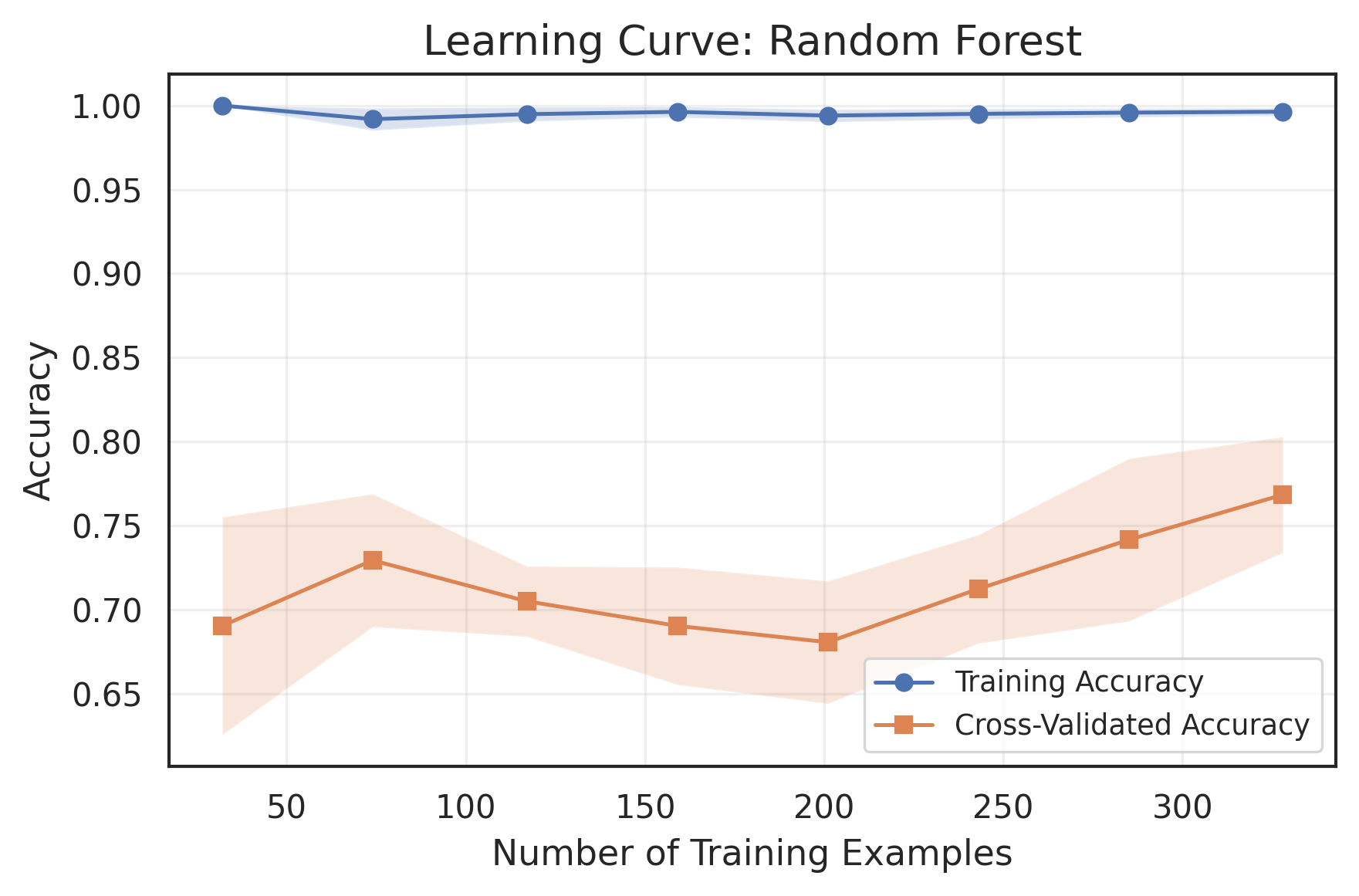}
    \caption{Learning curve for the RF model, showing training and cross-validated accuracy as a function of training-set size. Shaded regions denote ±1 standard deviation across cross-validation folds.}
    \label{fig:appendix_lc_rf}
\end{figure}

\begin{figure}[t]
    \centering
    \includegraphics[width=\linewidth]{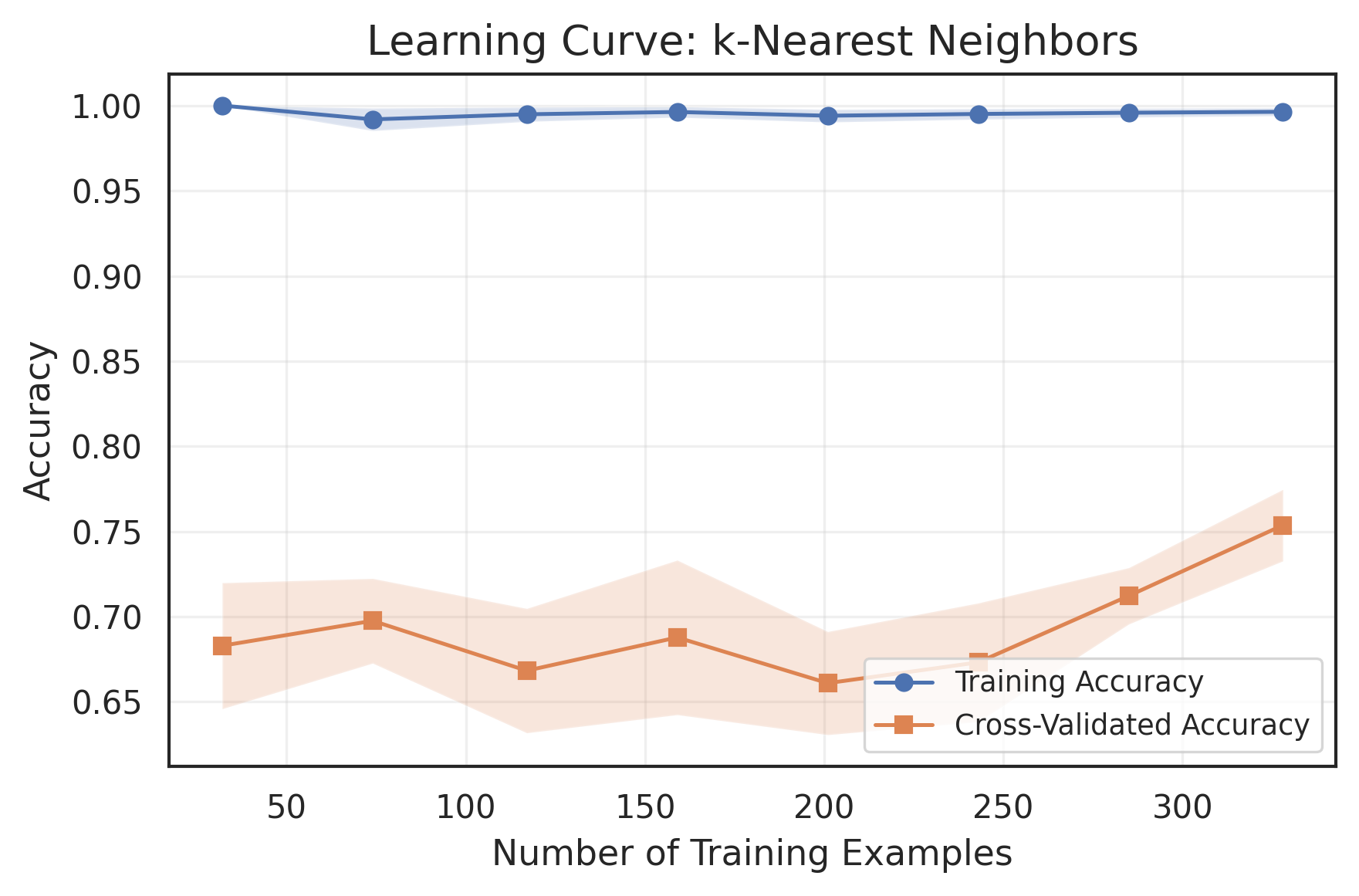}
    \caption{Learning curve for the KNN model, showing training and cross-validated accuracy as a function of training-set size. Shaded regions denote ±1 standard deviation across cross-validation folds.}
    \label{fig:appendix_lc_knn}
\end{figure}

\section{Random Subspace Robustness Analysis}
\label{app:subspace}

To evaluate sensitivity to feature redundancy and multicollinearity, we trained RF and KNN models on randomly sampled subsets of attitudinal predictors, using 50\%, 70\%, and 90\% of available features. For each configuration, models were trained and evaluated 50 times using the same fixed training and held-out test sets.

Table~\ref{tab:subspace_results} reports mean test accuracy, standard deviation, and observed ranges across random subspaces. Both models exhibit gradual performance degradation as fewer predictors are included, with no evidence of abrupt failure or dependence on a narrow subset of features. RF models demonstrate lower variance across subspaces than KNN, consistent with the stabilizing effects of ensemble learning.

\end{document}